\documentclass[12pt]{article}
\usepackage{graphicx}
\usepackage{epsfig}

\def\a{\alpha}
\def\b{\beta}

\def\f{\phi}

\def\h{\eta}

\def\m{\mu}
\def\n{\nu}

\def\p{\pi}
\def\q{\theta}
\def\r{\rho}
\def\s{\sigma}

\def\G{\Gamma}

\def\O{\Omega}

\def\Q{\Theta}

\def\ve{\varepsilon}

\def\beq{\begin{equation}}
\def\eeq{\end{equation}}
\def\bea{\begin{eqnarray}}
\def\eea{\end{eqnarray}}

\def\I{\mathrm{i}}

\def\pl#1#2#3{Phys.~Lett.~{\bf B {#1}} ({#2}) #3}
\def\np#1#2#3{Nucl.~Phys.~{\bf B {#1}} ({#2}) #3}
\def\prl#1#2#3{Phys.~Rev.~Lett.~{\bf #1} ({#2}) #3}
\def\pr#1#2#3{Phys.~Rev.~{\bf D {#1}} ({#2}) #3}

\def\prep#1#2#3{Phys.~Rep.~{\bf {#1}} ({#2}) #3}

\newcommand{\gl}{\tilde{g}}

\newcommand{\mpl}{M_{\rm P}}

\begin{document}
\date{\mbox{ }}
\title{{\normalsize DESY 03-078\hfill\mbox{}\\
July 2003\hfill\mbox{}}\\
\vspace{2cm} \textbf{Gauge Couplings at High Temperature\\ 
and the Relic Gravitino Abundance}\\
[8mm]}
\author{W.~Buchm\"uller, K.~Hamaguchi, M.~Ratz\\
\textit{Deutsches Elektronen-Synchrotron DESY, Hamburg, Germany}}
\maketitle

\thispagestyle{empty}

\begin{abstract}
\noindent
In higher-dimensional supersymmetric theories gauge couplings of the effective 
four-dimensional theory are determined by expectation values of scalar fields. We 
find that at temperatures above a critical temperature $T_*$, which depends on
the supersymmetry breaking mass scales, gauge couplings decrease like $T^{-\a}$,
$\a > 1$. This has important cosmological consequences. In particular it leads
to a relic gravitino density which becomes independent of the reheating temperature
for $T_R > T_*$. For small gravitino masses, $m_{3/2} \ll m_{\gl}$, the mass 
density of stable gravitinos is essentially determined by the gluino mass. The
observed value of cold dark matter, $\O_{\rm CDM}h^2 \sim 0.1$, is obtained for gluino
masses $m_{\gl} = {\cal O}(1 {\rm TeV})$.  
\end{abstract}

\newpage

In higher-dimensional supersymmetric theories \cite{gsw87}, where the standard model 
emerges as low-energy effective theory, 
gauge and Yukawa couplings are determined by expectation
values of gauge singlet `moduli' fields. In a cosmological context,
this implies that generically all couplings depend on the parameters of the
cosmological evolution, such as the Hubble parameter, temperature, or the 
cosmological constant.

In the following we study the dependence of gauge couplings on temperature. As we
shall see, this has important consequences for the production of gravitinos in the
early universe. `Vacuum alignment' at high temperatures causes a power-like
decrease of gauge couplings. This then leads to a relic gravitino density which 
becomes independent of the reheating temperature $T_R$ above a critical temperature 
$T_*$.

As a specific example, consider gaugino mediation \cite{kks00,clx00} which is an
attractive mechanism to generate a realistic mass spectrum of gauginos, higgsinos
and scalar quarks and leptons in the supersymmetric standard model. The source of
supersymmetry breaking is the vacuum expectation value of a gauge singlet chiral
superfield $S$, 
\begin{equation}\label{vac}
 \langle S\rangle = S_0 + \q\q F_S\;,
\end{equation} 
which is localized on a four-dimensional (4d) brane embedded in $D$-dimensional space
time. The coupling to bulk gauge fields, expressed in terms of 4d $N=1$ superfields,
is given by
\begin{eqnarray}\label{actionD}
 I_D & = & 
 \int d^4x\ d^{D-4}y\ d^2\q\ \left\{\frac{1}{4g_D^2}
	W^{a}W^{a}\right. \nonumber\\
 & & {}\hphantom{\int d^4x d^{D-4}y d^2\Q \left\{\right.}\left.
	{}+\delta^{(D-4)}(y-y_S)\frac{1}{4M}S\,W^{a}W^a
	+\dots\right\} + h.c.\;,
\end{eqnarray}
where $W^{a}$ is the supersymmetric field strength.
$M$ is a mass scale in the range between the compactification scale and the 
$D$-dimensional Planck mass,
\begin{equation}
{1\over V^{1/(D-4)}}  < M < M_D < \mpl\;.
\end{equation}
Here $V = \int d^{D-4}y$ is the volume of the compact dimensions, 
$M_D = (V M_D^{D-4})^{-1/2} \mpl$ and 
$\mpl = 1/(8\p G_N)^{1/2} = 2.4\times 10^{18}$~GeV is the 4d Planck mass.
For instance, with $1/V^{1/(D-4)} \simeq M_\mathrm{GUT} = 2\times 10^{16}$~GeV 
one obtains $M_D = 2\times 10^{17}$~GeV in the case $D=6$.

Inserting the expectation value (\ref{vac}) in the action (\ref{actionD}) one obtains 
for the 4d gauge coupling and for the gaugino mass,
\begin{eqnarray}
 \frac{V}{g_D^2} + \frac{\f_0}{M} &=& \frac{1}{g_0^2}\;, \label{coupling} \\
 m_{\gl} &=& {g_0^2\over 2}\frac{F_S}{M}\;, \label{mgluino}
\end{eqnarray}
where $\f_0 = {\rm Re}S_0$. For the SU(3)  gauge coupling of the standard model
one has $g_0^2(\m) \geq g_0^2(M_\mathrm{GUT}) \simeq {1\over 2}$. The gravitino mass
is given by 
\begin{equation}\label{mgrav}
 m_{3/2} = \h\frac{F_S}{\mpl}\;,
\end{equation}
where $\h \geq 1/\sqrt{3}$. The smallest gravitino mass is obtained if $F_S$ is the only 
source of supersymmetry breaking, which is the case in gaugino mediation. The gravitino
mass is then always smaller than the gaugino mass $m_{\gl}$,
\begin{equation}
 m_{\gl} \geq {g_0^2\over 2}\frac{F_S}{M_D} = 
{\sqrt{3}\over 2}g_0^2(V M_D^{D-4})^{1/2} m_{3/2} > m_{3/2}\;,
\end{equation}
since the volume enhancement factor $\r=(V M_D^{D-4})^{1/2}$ is larger than 
$2/(\sqrt{3} g_0^2) \leq 4/\sqrt{3}$. For instance, in $D=6$ one has $\r \sim 10$. 

The 4d effective
action for the zero modes contains a coupling of the scalar field $\f$ to the
supersymmetric gauge kinetic term, 
\begin{eqnarray}\label{eff}
 I_4 &=& 
 \int d^4x \left\{-\frac{1}{4}F^a_{\mu\nu}F^{a\mu\nu}-
 	 \I\lambda^a\sigma^{\mu}(D_{\mu}\bar{\lambda})^a 
     -{1\over 2} m_{\gl} \left(\lambda^a\lambda^a +  \bar{\lambda}^a\bar{\lambda}^a\right) 
\right.\nonumber\\
   & & {}\hphantom{\int d^4x \left\{\right.}\left.
       {} + g_0^2{\f\over M}\left(-\frac{1}{4}F^a_{\mu\nu}F^{a\mu\nu}-
 	\I\lambda^a\sigma^{\mu}(D_{\mu}\bar{\lambda})^a\right) + \dots\right\} \;;
\end{eqnarray}
here $F^a$ is the field strength of the vector potential $A^a$, and $\lambda^a$
denotes the gaugino.
At finite temperature the gauge kinetic term acquires an expectation value which leads 
to a force on the scalar field $\f$. This expectation value can be easily
calculated by making use of the anomalous divergence of the supercurrent \cite{gw85}, 
\begin{equation}\label{andiv}
 \bar{D}^{\dot\alpha}J_{\alpha\dot\alpha} = 
  \frac{1}{3}{\b(g_0)\over g_0} D_\alpha W^{a}W^a\;,
\end{equation}
which contains the trace anomaly of the energy momentum tensor,
\begin{equation}\label{trace}
 T^\mu{}_\mu =
 - 2 {\b(g_0)\over g_0}\left(-\frac{1}{4}F_a^{\mu\nu}F^{a\m\n}
 -\I\lambda^a\sigma^\mu\left(D_\mu \bar{\lambda}\right)^a\right)\;,
\end{equation}
where $\b(g_0)$ is the usual $\b$-function of the gauge coupling.

The thermal average of the energy momentum tensor is determined by energy density
and pressure,
\begin{equation}\label{thermtr}
\langle T^\mu{}_\mu \rangle_T = \ve - 3 P\;,
\end{equation}
which are related by
\begin{equation}\label{legen}
\ve = -P + T s = -P + T {\partial P\over \partial T}\;.
\end{equation}
The pressure has been calculated in perturbation theory for a gauge theory with 
fermions in the fundamental representation \cite{kap89}. Correcting for the colour
charge of the gauginos one obtains for a pure supersymmetric gauge theory,
\begin{equation}\label{press}
P = \left(a_0 - a_2 g_0^2(T) + \dots\right) T^4\;,
\end{equation}
with
\begin{equation}\label{a02}
a_0 = {\p^2\over 24} n_A\;, \quad a_2 = {1\over 64} T_A  n_A\;.
\end{equation}  
Here $T_A$ is the Dynkin index of the adjoint representation and $n_A = {\rm dim}\ G$,
i.e. the number of gluons. For SU($N$) one has $T_A = N$ and $n_A = N^2-1$.
From eqs.~(\ref{trace}) - (\ref{press}) one obtains for the thermal expectation
value of the gauge kinetic term,
\begin{equation}\label{thermkin}
\left\langle-\frac{1}{4}F^a_{\mu\nu}F^{a\mu\nu}
 -\I\lambda^a(\sigma^\mu D_\mu \bar{\lambda})^a\right\rangle_T = a_2 g_0^2 T^4 \;.
\end{equation}
Note that the sign of the expectation value is positive and that there is no
dependence on the $\b$-function. Because of the anomaly one no longer has $P=\ve/3$.

The mass of a chiral superfield, whose vacuum expectation value breaks supersymmetry, 
is generally controlled by the supersymmetry breaking mass scale, i.e.,
$m_\f \propto m_{3/2}$. Small fluctuations around
the minimum are then described by the lagrangian (cf.~(\ref{eff}), (\ref{thermkin})),
\begin{equation}\label{linear}
 {\cal L} = 
 {1\over 2}(\partial \f)^2 - {\xi\over 2} m_{3/2}^2 \f^2 + a_2 g_0^4 T^4 {\f\over M} \;.
\end{equation}
Hence, the thermal fluctuations of gauge bosons and gauginos induce a negative linear
term in the effective potential for $\f$. In many models the parameter $\xi$ is 
${\cal O}(1)$.

The negative linear term in the effective potential leads to an increase of the field
$\f$. Its equilibrium value at finite temperature is given by
\begin{equation}\label{ftherm}
\f_T = {a_2 g_0^4\over \xi} {T^4 \over m^2_{3/2} M}\;.
\end{equation}
Note that the fluctuations of $\f$ are not in thermal equilibrium and that $\f$ does not
acquire a thermal mass.
According to (\ref{coupling}) the shift in $\f$ changes the gauge coupling to 
$g(\f_T)$,
\begin{equation}\label{gtherm}
{1\over g_0^2} + {\f_T\over M} = {1\over g^2(\f_T)}\;.
\end{equation}
This change of the gauge coupling becomes significant at a temperature $T_*$ where
$\f_T/M \sim 1/g_0^2$, i.e.,
\begin{equation}
T_* = \left({\xi\over a_2 g_0^6}\right)^{1/4} \left(m_{3/2}M\right)^{1/2} \;. 
\end{equation}
Here we have assumed that $F_S$ does not depend on temperature, as in 
the Polonyi model.
Using eqs.~(\ref{mgluino}) and (\ref{mgrav}) the mass scale $M$ can be expressed in 
terms of gaugino and gravitino masses, which yields 
\begin{equation}\label{tcrit}
T_* = \left({\xi\over a_2 g_0^2 \eta^2}\right)^{1/4}
\left({m^2_{3/2}\mpl\over 2 m_{\gl}}\right)^{1/2} \;.
\end{equation}
Extrapolating eqs.~(\ref{ftherm}) and (\ref{gtherm}) to temperatures larger than $T_*$
leads to a rapid decrease of the gauge coupling as $g^2(\f_T) \propto 1/T^4$.

However, at large values of $\f_T/M$ the effective lagrangian (\ref{linear}) is
no longer appropriate. First, the decrease of the gauge coupling reduces the force of
the thermal bath on the field $\f$. This backreaction can be taken into account by
using as effective potential the free energy density of the thermal system evaluated
with the field-dependent gauge coupling,
\begin{equation}
f = -P = \left(-a_0 + a_2 g^2(T,\f) + \ldots \right)\;,
\end{equation}
where $g(T,\f)$ has to be determined from the equations of motion.
Second, for large values of $\f$, higher powers of $\f/M$ have to be taken into account.
This leads to the effective lagrangian
\begin{equation}\label{nonlinear}
 \bar{{\cal L}} = 
 {1\over 2}(\partial \f)^2 - {1\over 2} m_{3/2}^2 h(\f) - a_2 g^2(T,\f) T^4 \;,
\end{equation}
where
\begin{equation}
g^2(T,\f) = {g_0^2(T)\over 1 + g_0^2(T) k(\f)}\;, 
\end{equation}
\begin{equation}
h(\f) = \xi \f^2 \left(1 + {\cal O}\left({\f\over M}\right)\right)\;,\quad
k(\f) = {\f\over M}\left(1 + {\cal O}\left({\f\over M}\right)\right)\;.
\end{equation} 
$k(\f)$ replaces the linear term $\f/M$ in eq.~(\ref{eff}).
The equilibrium value of $\f$ is now determined by the equation
\begin{equation}
{h'(\f_T)(1+g_0^2 k(\f_T))^2\over k'(\f_T)} = 2 a_2 g_0^4 {T^4\over m^2_{3/2}} \;.
\end{equation} 
For small values of $\f$ one recovers eq.~(\ref{ftherm}). Neglecting corrections
${\cal O}(\f/M)$ for $h(\f)$ and $k(\f)$, keeping only the effect of the back
reaction, one obtains at large temperatures $\f_T \propto T^{4/3}$ and correspondingly
for the gauge coupling $g^2(T,\f_T) \propto T^{-4/3}$. This decrease with temperature
is much weaker than the $T^{-4}$ fall-off obtained in the linear approximation. We expect
that the true decrease, which is determined by the back reaction together with the
behaviour of $h$ and $k$ at large values of $\f$, lies somewhere in between. 

The time evolution of the field $\f$ is determined by the equation of motion
\begin{equation}
 \ddot{\f} + 3H \dot\f + {1\over 2} m_{3/2}^2 h'(\f)
-{a_2 g_0^4 \over \left(1+g_0^2 k(\f)\right)^2} k'(\f) T^4  = 0\;,
\end{equation}
where $H$ is the Hubble parameter. For $H > m_{3/2}$ the motion is damped whereas for
$H < m_{3/2}$ the field $\f$ oscillates. During the period of reheating the Hubble
parameter generally depends not only on the thermal bath, but also on the time evolution
of other fields, in particular the inflaton. The detailed analysis of the time evolution
of $\f$ is beyond the scope of this paper. In the following we shall assume that at the
end of reheating thermal equilibrium is achieved and that, to good approximation, $\f$
is close to its equilibrium value $\f_T$. 

The power-like fall-off of gauge couplings at high temperature, 
$g^2 \propto T^{-\a}$ with $\a > 1$, has important cosmological implications. 
An immediate consequence is that one loses thermal equilibrium at a temperature
$T_{\rm eq}$ much below the unification scale $M_{\rm GUT}$.
For instance, for $\a = 2$, $m_{\gl} \simeq 1$~TeV and $m_{3/2} \simeq 100$~GeV,  one 
obtains $\G(T_{\rm eq}) \simeq H(T_{\rm eq})$ at 
$T_{\rm eq} \sim (m^2_{3/2}M_\mathrm{P}^2/m_{\gl})^{1/3} \sim 10^{12}$~GeV.  
The decrease of the gauge coupling also crucially affects the production of
gravitinos after inflation \cite{kl84} which we now discuss.

The thermal production of gravitinos by gluons, gluinos, quarks and squarks is governed 
by the Boltzmann equations. The collision term has been calculated to leading order in
the gauge coupling. For the gauge group SU($N$), with $2n_f$ chiral multiplets in the
fundamental representation, one has \cite{bbb01}, 
\begin{eqnarray}\label{boltz}
  \frac{dn_{3/2}}{dt}+3Hn_{3/2} &=& C_{3/2}(T,\phi)
  \nonumber\\
  &=& \frac{3\zeta (3)}{32\pi^3}
  g^2 (N^2-1)\frac{T^6}{\mpl^2}
  \left(1+\frac{m_{\gl}^2}{3m_{3/2}^2}\right)
       {\cal F}(T)\;,
\end{eqnarray}
where 
\begin{equation}
  {\cal F}(T) =
  \left(\ln\left(\frac{T^2}{m_{\rm gluon}^2(T)}\right)+0.3224\right)
  (N+n_f) + 0.5781n_f \;,
\end{equation}
with the thermal gluon mass 
\begin{equation}
  m_{\rm gluon}^2(T) = \frac{g^2}{6}(N+n_f)T^2 \;.
\end{equation}
For the gauge coupling we use $g(T,\f_T)$, except in case of the gluon mass which 
enters only logarithmically.

In the supersymmetric standard model gravitino production is dominated by QCD,
the strong interactions, where we have $N=3$ and $n_f=6$. If the gravitino is the LSP
and the GUT relations for gaugino masses hold, one has $m_{3/2} \ll m_{\gl}$.
Integrating eq.~(\ref{boltz}) up to a reheating temperature $T_R > T_*$, assuming 
a power decrease of the gauge coupling,
\begin{equation}
g^2(T,\f_T) \simeq \frac{g_0^2(T)}{1+(T/T_*)^\a}\;,
\end{equation}
one obtains a number density to entropy density ratio of gravitinos which is independent
of $T_R$,
\begin{eqnarray}
\left.{n_{3/2} \over s}\right|_{T_0} = {C_{3/2}(T_*,0)\over s(T_*)H(T_*)}I_{(\a)}\;.
\end{eqnarray}
Here $T_0$ is the present temperature,
$s = (2\pi^2/45)g_*(T)T^3$ is the entropy density, with $g_*(T_*)=915/4$ 
in the supersymmetric standard model, and
\begin{equation}\label{inta}
I_{(\a)}  =  \int^{\infty}_0\frac{dz}{(1+z^\a)^3} = 0.50 \ldots 0.73 \;,
\end{equation}
for $\a = 1 \dots 4$.
Inserting the expression for the collision term in 
eq.~(\ref{boltz}) one finds for the energy density to entropy density ratio of 
gravitinos ($T_R > T_*$),
\begin{eqnarray}\label{first}
  \left.\frac{\rho_{3/2}}{s}\right|_{T_0}
  &=&    
  \frac{135\sqrt{10}\zeta(3)}{64\pi^6}
  \frac{N^2-1}{g_*^{3/2}(T_*)}    
  \frac{T_* m_{\gl}^2(T_*)}{M_\mathrm{P} m_{3/2}}
  I_{(\a)} g_0^2(T_*){\cal F}(T_*)\;.
\end{eqnarray}
At temperatures $T_R$ much larger than $T_*$ also contributions involving Yukawa
interactions may become important, which remains to be studied.

\begin{figure}
\centerline{\psfig{figure=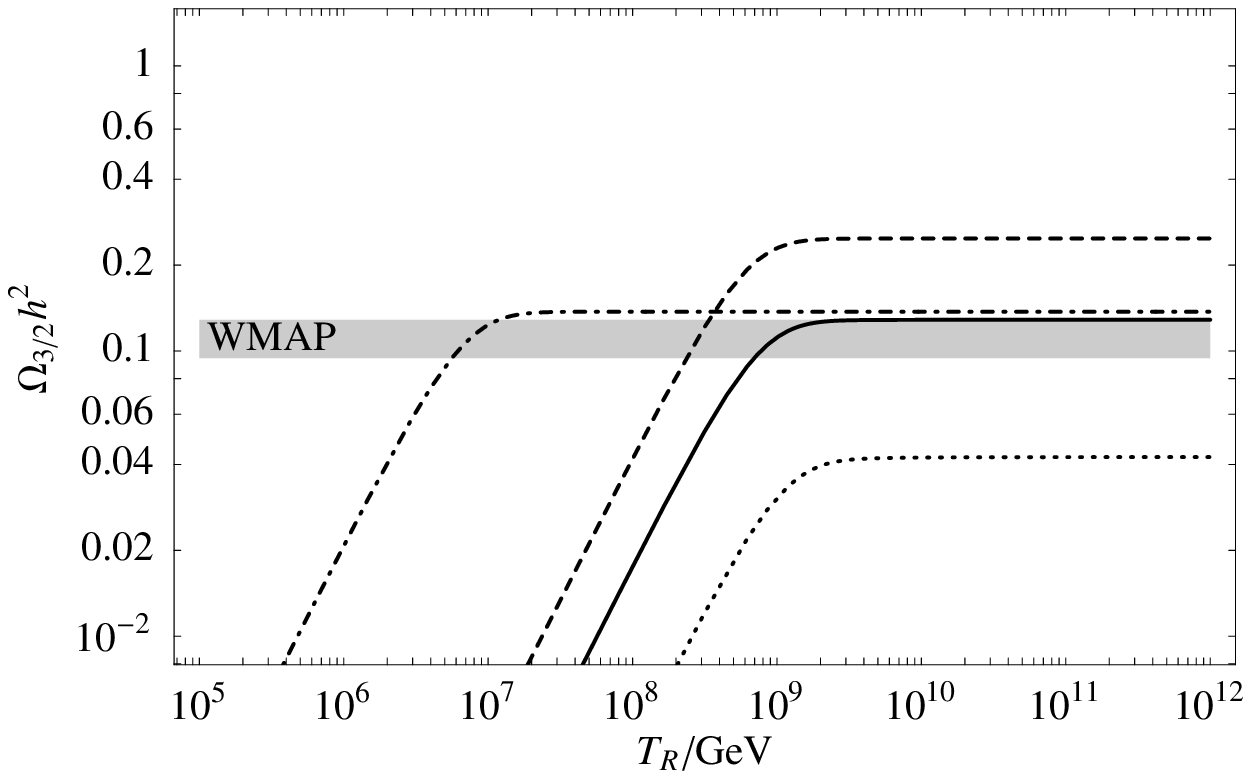,width=11cm}}
\caption{Relic gravitino density $\O_{3/2}h^2$ as function of the reheating temperature
$T_R$ for different gravitino and gluino masses: $m_{3/2} = 20$~GeV with 
$m_{\gl} = 1.5$~TeV (dashed line), $m_{\gl} = 1.0$~TeV (full line), 
$m_{\gl} = 0.5$~TeV (dotted line), and $m_{3/2} = 200$~MeV with 
$m_{\gl} = 1.0$~TeV (dashed-dotted line); $\xi/\eta^2 =1$, $\a = 2$.
$\O_{3/2}h^2$ reaches a plateau at $T_R \simeq T_* \propto m_{3/2}/\sqrt{m_{\gl}}$. 
The band denotes the WMAP result for cold dark matter with a $2\sigma$ error.}
\vspace{1.5cm}
%\end{figure}
%\begin{figure}
\centerline{\psfig{figure=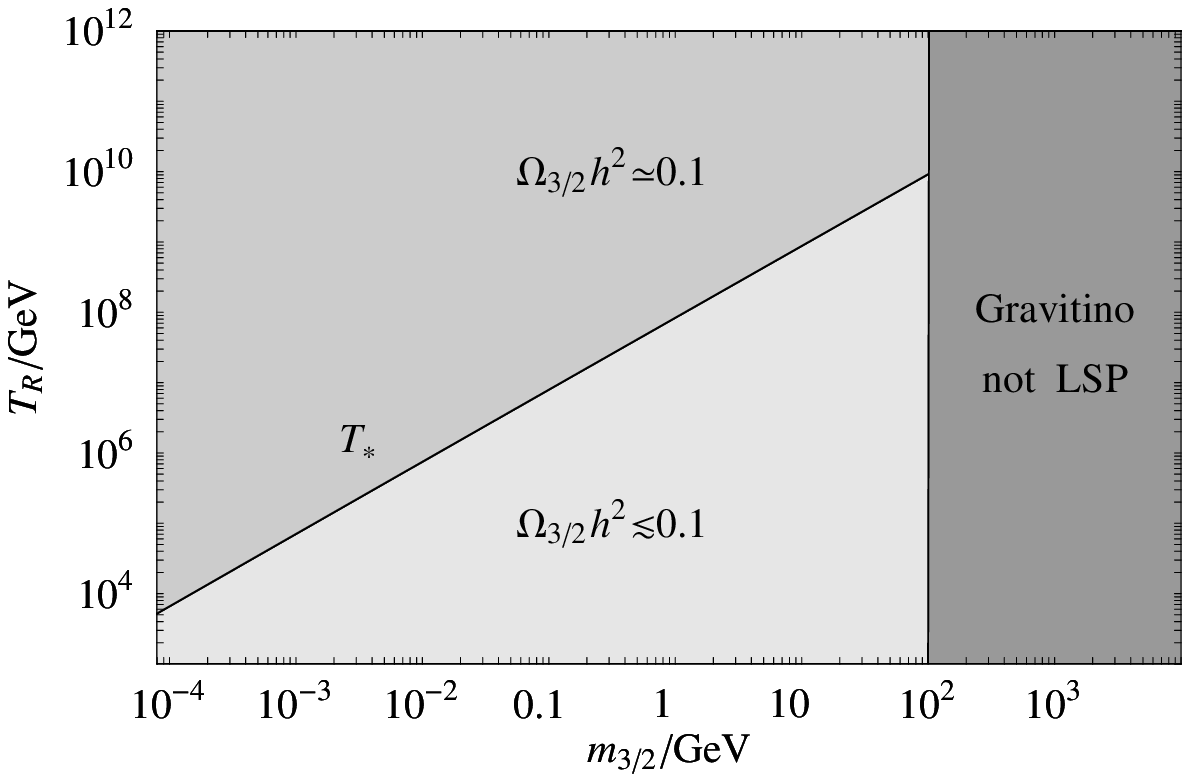,width=11cm}}
\caption{Relic gravitino density for different values of reheating temperature and
gravitino mass. $\xi/\eta^2 =1$. $m_{\gl}=1$~TeV, which implies
$m_{3/2} < 0.1$~TeV for a stable gravitino. For $T_R > T_*$, $\O_{3/2}h^2$ is independent
of $T_R$ and $m_{3/2}$. }
\end{figure}

One can now insert the relation (\ref{tcrit}) between the temperature $T_*$ and gluino 
and gravitino masses into eq.~(\ref{first}), which yields the result ($T_R > T_*$),
\begin{eqnarray}
  \left.\frac{\rho_{3/2}}{s}\right|_{T_0} =
  \frac{135\sqrt{5}\zeta(3)}{64\pi^6}
  \frac{N^2-1}{g_*^{3/2}(T_*)}
  \left({m_{\gl}^{3/2}(\mu)\over \mpl^{1/2}}\right)
  \left(\frac{\xi}{a_2\eta^2}\right)^{1/4}
  I_{(\a)}\hat{{\cal F}}(T_*)\;.
\end{eqnarray}
Here we have used the gluino mass at a scale $\mu$ as parameter,
and $\hat{{\cal F}}(T_*) = {\cal F}(T_*) g_0^{9/2}(T_*)/g_0^3(\mu)$ is
a factor ${\cal O}(1)$ which takes gauge couplings and their running into account.
Remarkably, in $\rho_{3/2}/s$ the dependence on the gravitino mass has dropped out.
For the dominant QCD contribution $N=3$ and $a_2 = 3/8$ (cf.~(\ref{a02})). 
Dividing by the critical density
$\rho_{\rm crit}/s = 3.65h^2\times 10^{-9}$~GeV \cite{rpp02} one finally obtains 
($T_R > T_*$),
\begin{eqnarray}\label{final}
  \Omega_{3/2}h^2 =  0.1 \times
  \left(\frac{m_{\gl}(1~{\rm TeV})}{1.0~{\rm TeV}}\right)^{3/2}
  \left(\frac{\xi}{\eta^2}\right)^{1/4}
  I_{(\a)}\hat{{\cal F}}(T_*)\,.
\end{eqnarray}
For gaugino mediation one has $\xi/\eta^2 = {\cal O}(1)$; in the temperature range
$T_* = 10^4 \cdots 10^{12}$~GeV we estimate 
$I_{(\a)}\hat{{\cal F}}(T_*) = 0.5\ldots 2$. 
It is then very
astonishing how close the obtained value for $\Omega_{3/2}h^2$ is to the observed
one for cold dark matter for gluino masses ${\cal O}(1~{\rm TeV})$. The WMAP
collaboration recently obtained ($2\s$ error),
$\O_{{\rm CDM}} h^2 = (\O_m - \O_b)h^2 = 0.113^{+0.016}_{-0.018}$ \cite{wmap}.
The relic gravitino density $\O_{3/2}h^2$ is shown in fig.~1 as function of the
reheating temperature $T_R$ for different values of $m_{\gl}$ and $m_{3/2}$. At
$T_R \simeq T_*$  the density reaches a plateau whose value is 
essentially independent of $T_R$ and $m_{3/2}$. The figure clearly shows the scaling 
$T_* \propto m_{3/2}/\sqrt{m_{\gl}}$. 

One may also use eq.~(\ref{final}) to determine the range of gluino masses consistent with
the WMAP result for cold dark matter. Varying $\O_{\rm CDM} h^2$ and 
$I_{(\a)} \hat{{\cal F}}(T_*)$ in the ranges specified above we find,
\begin{equation}\label{rangemgl}
m_{\gl}\ =\ (0.5 \ldots 2.0)\ {\rm TeV} \left({\h^2\over \xi}\right)^{1/6}\;.
\end{equation}
Hence, the hypothesis that gravitinos are the dominant component of dark matter will
be tested at LHC\ !

The range for the gluino mass given in eq.~(\ref{rangemgl}) has been obtained in
the case of gaugino mediation where $m_{3/2} = (2\eta/g_0^2)\,(M/M_\mathrm{P})\,
m_{\gl}$ (cf.\ (\ref{mgluino}) and (\ref{mgrav})), with $\eta={\cal O}(1)$. For
gravity mediation \cite{nilles}, one obtains the same results with $\eta$ 
replaced by $\eta'=\eta\,M_\mathrm{P}/M$. $m_{3/2}$ and $m_{\gl}$ now have the 
same order of magnitude, but the gravitino can be the LSP without fine tuning. 
The range for the gluino mass remains unchanged unless $M$ is smaller than 
$M_\mathrm{P}$ by several orders of magnitude. In the case of gauge mediation 
\cite{giudicerattazzi}, $\eta$ has to be replaced by $\eta'=\eta\,8\pi^2\langle
X\rangle/M$ where $\langle X\rangle$ is the messenger scale. The mass range 
(\ref{rangemgl}) for the gluino mass is then obtained if $M$ is of order 
$8\pi^2\,\langle X\rangle$. Note that the rapid decrease of gauge couplings at 
high temperature occurs independently of the supersymmetry breaking mechanism.

Our results have important consequences for leptogenesis \cite{fy86} where
the typical baryogenesis temperature is $T_B = {\cal O}(10^{10}~{\rm GeV})$ or larger
\cite{bdp02}. According to previous studies this implies that unstable gravitinos 
have to be heavier than a few TeV \cite{kkm01,cex03}. Stable gravitinos may have 
masses below ${\cal O}(1~{\rm keV})$ \cite{pp82} so that their mass density is below 
the critical density even when they are thermalized. Further, it has been shown that 
also gravitino masses $m_{3/2} \sim 10\ldots 100~{\rm GeV}$ can be consistent, which 
then constrains masses and couplings of other neutralinos and sleptons 
\cite{bbp98,ahs00}. 

Our analysis shows that there is no constraint on the reheating temperature for 
gluino masses below ${\cal O}(1~{\rm TeV})$ (cf.~fig.~2) if the gravitino is the
lightest supersymmetric particle. For 
$m_{\gl} = {\cal O}(1~{\rm TeV})$ and reheating temperatures $T_R > T_*$ we find
$\O_{3/2} h^2 \simeq \O_{\rm CDM} h^2 \simeq 0.1$, independently of $m_{3/2}$.

The maximal value of the critical temperature $T_*$ 
is obtained for $M\sim \mpl$ and $m_{3/2} \sim 100$~GeV, so that the gravitino can still 
be the LSP for a gluino mass ${\cal O}(1~{\rm TeV})$. This yields 
$T_*^{\rm max} \sim 10^{10}$~GeV, which happens to coincide with the typical leptogenesis 
temperature. Hence, for a reheating temperature $T_R$ larger than the leptogenesis 
temperature $T_B$, relic gravitinos always have the observed dark matter energy 
density $\O_{\rm CDM} h^2$ if the gluino mass is ${\cal O}(1~{\rm TeV})$. In this way 
the supersymmetry breaking scale in the observable sector is directly determined by the 
dark matter density $\O_{\rm CDM} h^2$, 
independently of the supersymmetry breaking scale in the hidden sector!

The interplay of particle physics and cosmology relates some properties of the 
universe to properties of elementary particles. Of particular interest is the composition
of the present energy density $\O h^2$. Leptogenesis explains the baryon density 
$\O_b h^2$ in terms of neutrino masses and mixings. 
As we have seen, for stable gravitinos the dark matter 
density $\O_{\rm CDM} h^2$ is then determined by the gluino mass, i.e., the supersymmetry 
breaking scale in the observable sector, which may also be responsible for the dominant
contribution to $\O h^2$, the cosmological constant.

\vspace{1cm}
\noindent
We are grateful to D.~B\"odeker, A.~Brandenburg, L.~Covi, M.~B.~Einhorn and 
A.~Hebecker for helpful comments and discussions. Two of us (W.B., M.R.) would  
like to thank G.~Kane and the Michigan Center for Theoretical Physics for 
hospitality during the {\it Baryogenesis Workshop} where part of this work has 
been carried out. K.H. thanks the Japan Society for the Promotion of Science 
for financial support.

%\newpage

\end{document}